\documentclass[notitlepage,twocolumn,letterpaper,natbib,aps,prl,amsmath,amsfonts,nofootinbib,preprintnumbers,superscriptaddress,secnumarabic,groupedaddress]{revtex4-1}
\pdfoutput=1
\usepackage{amssymb,amsmath,latexsym,mathrsfs}
\usepackage{url}
\usepackage{enumitem}
\usepackage{graphicx}
\usepackage[usenames,dvipsnames]{color}
\usepackage[breaklinks,colorlinks,urlcolor=blue,citecolor=blue,linkcolor=magenta]{hyperref}

\usepackage{multirow}
\usepackage{float}
\usepackage{cases}
\usepackage{blindtext}
\usepackage{hhline}

\makeatletter
\renewcommand\tableofcontents{%
    \@starttoc{toc}%
}
\makeatother

\usepackage{titlesec}

\titleformat{\section}
  {\normalfont\fontsize{12}{12}\bfseries}{\centering \thesection}{1em}{}

\titleformat{\subsection}
  {\normalfont\fontsize{11}{11}\bfseries}{\centering \thesubsection}{0.5em}{}

\titlespacing*{\subsection}
  {0pt}{1\baselineskip}{0.5\baselineskip}

\titlespacing*{\section}
  {0pt}{1.6\baselineskip}{0.9\baselineskip}

\titleformat{\subsubsection}
  {\normalfont\fontsize{10}{10}\bfseries}{\centering \thesubsubsection}{1em}{}
\titlespacing*{\subsection}
  {0pt}{0.9\baselineskip}{0.4\baselineskip}
  
\usepackage{dsfont}

\begin{document}
\title{Improved BBN Constraints on the Variation of the Gravitational Constant}

\preprint{KCL-2019-81}

\author{James Alvey$^\mathds{A}$}

\author{Nashwan Sabti$^\mathds{S}$}

\author{Miguel Escudero$^\mathds{E}$}

\author{Malcolm Fairbairn$^\mathds{F}$}

\affiliation{\vspace{8pt}Theoretical Particle Physics and Cosmology Group\\ 
King's College London, Department of Physics, Strand, London WC2R 2LS, UK}

\def\thefootnote{$\mathds{A}$\hspace{-0.5pt}}\footnotetext{\href{mailto:james.alvey@kcl.ac.uk}{james.alvey@kcl.ac.uk}}
\def\thefootnote{$\mathds{S}$\hspace{0.4pt}}\footnotetext{\href{mailto:nashwan.sabti@kcl.ac.uk}{nashwan.sabti@kcl.ac.uk}}
\def\thefootnote{$\mathds{E}$}\footnotetext{\href{mailto:miguel.escudero@kcl.ac.uk}{miguel.escudero@kcl.ac.uk}}
\def\thefootnote{$\mathds{F}$}\footnotetext{\href{mailto:malcolm.fairbairn@kcl.ac.uk}{malcolm.fairbairn@kcl.ac.uk}}
\setcounter{footnote}{0}
\def\thefootnote{\arabic{footnote}}
\begin{abstract}
\noindent Big Bang Nucleosynthesis (BBN) is very sensitive to the cosmological expansion rate. If the gravitational constant $G$ took a different value during the nucleosynthesis epoch than today, the primordial abundances of light elements would be affected. In this work, we improve the bounds on this variation using recent determinations of the primordial element abundances, updated nuclear and weak reaction rates and observations of the Cosmic Microwave Background (CMB). When combining the measured abundances and the baryon density from CMB observations by Planck, we find $G_\mathrm{BBN}/G_0 = 0.99^{+0.06}_{-0.05}$ at $2\sigma$ confidence level. If the variation of $G$ is linear in time, we find $\dot{G}/G_0 = 0.7^{+3.8}_{-4.3}\times 10^{-12} \, \mathrm{yr}^{-1}$, again at $2\sigma$. These bounds are significantly stronger than those from previous primordial nucleosynthesis studies, and are comparable and complementary to CMB, stellar, solar system, lunar laser ranging, pulsar timing and gravitational wave constraints.
\end{abstract}

\maketitle

\emph{Introduction.}--- \setlength\parskip{0pt}
Ever since the Large Number Hypothesis of Dirac, physicists have wondered whether the constants of nature may evolve over time \cite{1937Natur.139..323D}.  Since then the mathematical framework for such a variation has been developed. This was done first with ideas such as Kaluza-Klein theory where the coupling between gravity and matter sectors was set by the size of a compact dimension \cite{Kaluza:1984ws,Klein:1926tv}, and then by advancements such as Brans-Dicke theory where the coupling between matter and gravity is endowed with dynamics \cite{Brans:1961sx}.  A key problem in the ongoing attempt to unify the four forces of nature using string theory is figuring out how to stabilise the higher dimensions, and consequently the value of the low energy couplings, including the gravitational constant $G$ \cite{Douglas:2006es}.  

A theory where $G$ is allowed to vary almost always involves (by definition) promoting $G$ to be related to the expectation value of some dynamical scalar field. For consistency, the scalar field will then have a kinetic term and a potential.  In order to observe any dynamical changes of such a field over cosmological time, the curvature of that potential, and therefore the mass of this field, has to be very small. Constraints from solar system observations, like radar ranging of Mars, place very tight limits on the contribution to gravity due to such light scalars in the Universe today \cite{Bertotti:2003rm}.\footnote{Notable exceptions to this are Chameleon type theories where the mass of such scalar fields can be affected by the local density of matter \cite{Khoury:2003aq}.}

On the other hand, we are in possession of much less information about the detailed behaviour of gravity in the early Universe, in particular during the period of Big Bang Nucleosynthesis at which point the Universe is expected to be extremely uniform.  Most scenarios where the value of $G$ is set by a scalar field obtaining an expectation value result in an effective mass for that field that is much larger than the energy scales relevant even at this earlier epoch. As such, we do not have any good reason to {\it expect} that $G$ is likely to be different.  Nevertheless, if we have the technology to place new limits on the variation of a fundamental constant like this one, we should take the opportunity to obtain those limits to the best of our abilities.  The subject may also become more important in the coming years given upcoming atomic interferometry experiments which will test extensions of gravity \cite{Biedermann:2014jya}. 

Strictly speaking, since $G$ is dimensionful, we should be very careful when we discuss its time variation. This is because any physically meaningful change in couplings should be written as a change in the dimensionless ratio between two dimensionful quantities.  We are assuming a scenario where $G=M_{\mathrm{Pl}}^{-2}$ changes while none of the particle physics energy scales vary.  For example, consider the case where all the dimensionful parameters of the Standard Model (like the Brout-Englert-Higgs expectation value $v$ and $\Lambda_\mathrm{QCD}$) are determined by dimensional transmutation of couplings set at a single energy scale $M_\mathrm{GUT}$. If gravitational corrections to the running of those couplings are irrelevant, then discussing variations in $M_\mathrm{Pl}/M_\mathrm{GUT}$ becomes better defined \cite{Duff:2014mva}.  This is the kind of situation we are considering.  In what follows, we will refer to $G$ as the value of Newton's gravitational constant over time and $G_0$ as its value as measured today.

Big Bang Nucleosynthesis (BBN) is sensitive to modifications of the expansion history in the early Universe. Indeed, predictions for the primordial element abundances are strongly dependent on the Hubble rate $H$ and subsequently on the value of the gravitational constant $G$ during the relevant epoch. Recent improvements in measurements of these early Universe abundances ~\cite{2017RMxAC..49..181P,Aver:2015iza,Izotov:2014fga, Cooke:2016rky,Balashev:2015hoe,2018MNRAS.477.5536Z,Riemer-Sorensen:2017pey} allow us to improve constraints on the gravitational physics during the BBN era. With some mild assumptions detailed in the \emph{Methodology} section, this can be translated into bounds on the variation in the value of $G$ between nucleosynthesis and today.

Nucleosynthesis is of course not the only scenario in which changes in $G$ would be manifest. Bounds on the time variation of $G$ can be inferred from Cosmic Microwave Background (CMB) \cite{Bai:2015vca, Ooba:2017gyn}, stellar \cite{GarciaBerro:2011wc, Mould:2014iga, Bellinger:2019lnl}, solar system \cite{Chandler:1993, Pitjeva:2013chs, Fienga:2014bvy}, pulsar timing \cite{Kaspi:1994hp, Zhu:2018etc}, lunar laser ranging \cite{Williams:2004qba, Hofmann:2018myc} and gravitational wave \cite{Yunes:2009bv, Zhao:2018gwk} measurements. These place constraints at the same order of magnitude as those from BBN \cite{Campbell:1994bf,Dent:2001ga,Copi:2003xd, Clifton:2005xr}, with the strongest bound being $\dot{G}/G_0 = (7.1 \, \pm \, 7.6) \times 10^{-14} \, \mathrm{yr}^{-1}$ at the 68\% CL from lunar laser ranging experiments \cite{Hofmann:2018myc}. Note that each given constraint probes different epochs in the evolution of the Universe, and should be compared as such. For comprehensive reviews of constraints on $G$ and newer developments, see \cite{Uzan:2002vq, Uzan:2010pm, Martins:2017yxk}.

It is the nucleosynthesis bounds which we look to update in this work. We do this by \emph{i)} using up-to-date measurements of the primordial element abundances reported by the PDG \cite{pdg}, \emph{ii)} including a weak determination of the baryon density \cite{Bai:2015vca} to alleviate the degeneracy between $\Omega_{\mathrm{b}}h^2$ and $G$ in the deuterium abundance, \emph{iii)} accurately accounting for incomplete neutrino decoupling following \cite{Escudero:2018mvt,Escudero:2019new}, and \emph{iv)} making use of the state-of-the-art BBN code \texttt{PRIMAT} \cite{Pitrou:2018cgg} which has updated nuclear reaction rates and accounts for many corrections to the weak reaction rates. The methodology applied here follows that in \cite{Sabti:2019mhn}. 
\setlength\parskip{6pt}

\begin{figure}[t]
    \centering
    \includegraphics[width=\linewidth]{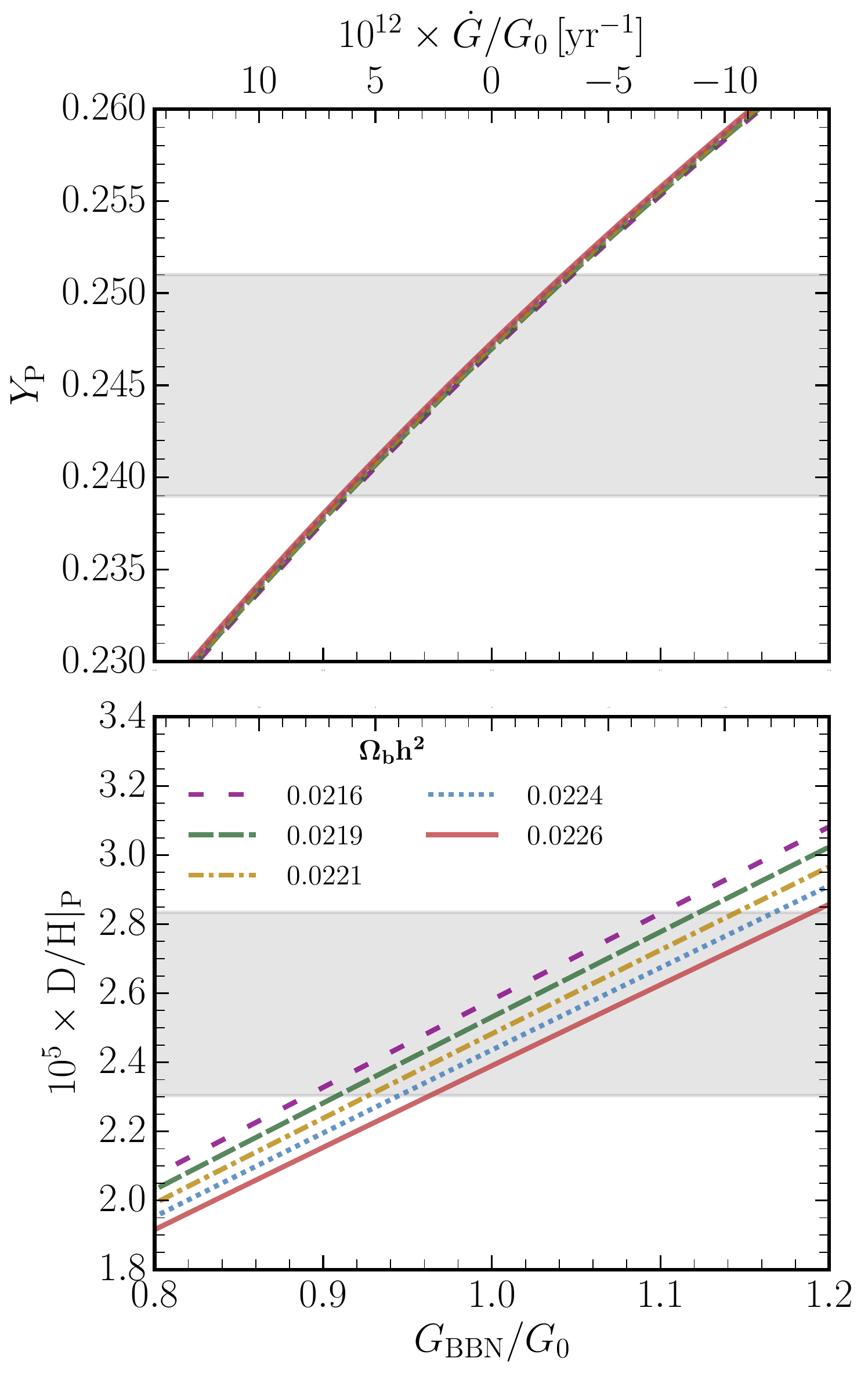}\vspace{-0.3cm}
    \caption{The variation in the primordial helium abundance (\emph{upper}) and the primordial deuterium abundance (\emph{lower}) as a function of $G_{\mathrm{BBN}}/G_0$. The grey bands correspond to the $2\sigma$ confidence intervals from the astrophysical measurements in Eqs. \eqref{eq:Yp} and \eqref{eq:DH}. The different lines correspond to representative values of the baryon density $\Omega_{\mathrm{b}}h^2$.}
    \label{fig:abundances}
\end{figure}

\emph{Cosmological Implications.}--- We can understand the effect on the primordial abundances of a different value of $G$ during nucleosynthesis, $G_{\mathrm{BBN}} \neq G_0$, in terms of the Hubble expansion rate $H \propto \sqrt{G}$. The dominant effect of an increased (decreased) expansion rate is to alter the time at which various weak and nuclear processes freeze-out. In particular, the proton-to-neutron conversion processes and the $p + n \leftrightarrow D + \gamma$ reaction will freeze-out earlier (later). This leads to an over(under)-production of both helium and deuterium compared to the case where $G_{\mathrm{BBN}} = G_0$. This is seen clearly in Figure \ref{fig:abundances}. From just these figures and the errors in the measurements of the abundances (indicated by the grey bands), we expect that our analysis will be able to constrain variations in $G$ below the 10\% level. 

\emph{Methodology.}--- To derive the bounds on the variation of the gravitational constant, we follow the approach presented in \cite{Sabti:2019mhn}. We make use of the publicly available code \texttt{NUDEC\_BSM}~\cite{Escudero:2018mvt, Escudero:2019new} to compute the background cosmology, including the effects of non-instantaneous neutrino decoupling. The relevant cosmological parameters are subsequently forwarded to the state-of-the-art BBN code \texttt{PRIMAT}~\cite{Pitrou:2018cgg}, which takes care of the nuclear reaction network and time evolution of primordial abundances. In all simulations we take the neutron lifetime to be the default value in \texttt{PRIMAT}: $\tau_n = 879.5 \, \mathrm{s}$. The abundances scale with $G/G_0$ in a way that closely matches the semi-analytic relations given in \cite{Fields:2019pfx}.
\setlength\parskip{0pt}

To quantify the effect of a time variation in $G$ we choose to parametrise the evolution as a slowly evolving linear function of time $t$ \cite{Uzan:2002vq},
\begin{align}
    G(t) = G_{\mathrm{BBN}} + \dot{G} \times (t - t_{\mathrm{BBN}}), \label{eq:gt}
\end{align}
so that $G_0 = G_{\mathrm{BBN}} + \dot{G}(t_0 - t_{\mathrm{BBN}})$. Here $t_0$ is the current age of the Universe such that $(t_0 - t_{\mathrm{BBN}}) \simeq 13.8\, \mathrm{Gyr}$~\cite{Aghanim:2018eyx}. This slowly varying function ensures that the assumption of a constant $G$ during the cosmologically very short period of primordial nucleosynthesis is an excellent approximation. In the \emph{Results} section, we will quote bounds on both $G_{\mathrm{BBN}}/G_0$ and $\dot{G}/G_0$, where the latter can be derived from Equation~\eqref{eq:gt}.
\setlength\parskip{6pt}

\emph{Data Analysis.}--- We make use of the observed primordial abundances of helium and deuterium as reported by the PDG \cite{pdg}. At 68\% CL, these are:

\vspace*{-0.55 cm}
\begin{align}
Y_{\rm P} 	            &= 0.245 \pm 0.003 \label{eq:Yp}\, , \\
{\rm D/H}|_{\rm P}   	&= (2.569  \pm 0.027) \times 10^{-5} \label{eq:DH} \,.
\end{align}
We also include theoretical errors in the predictions of $Y_{\rm P}$ and ${\rm D/H}|_{\rm P}$ due to uncertainties in the various nuclear reaction rates and the neutron lifetime \cite{Fields:2019pfx};
\begin{align}
\sigma(Y_{\rm P})^{\rm Theo} 	        		&= 0.00018 \,, \label{eq:sigmaYP} \\
\sigma({\rm D/H}|_{\rm P} )^{\rm Theo}   	&= 0.13 \times 10^{-5} \label{eq:sigmaDH} \, .
\end{align}
\setlength\parskip{0pt}

To quantify deviations from the measured primordial abundances due to changes in the gravitational constant, we construct a $\chi^2$ for BBN as follows:
\begin{align}\label{eq:chiBBN}
\chi_{\rm BBN}^2 =\, &\frac{\left[Y_{\rm P} - Y_{\rm P}^{\rm Obs}\right]^2}{\sigma^2(Y_{\rm P})^{\rm Theo} + \sigma^2(Y_{\rm P})^{\rm Obs}}\nonumber\\
&+ \frac{\left[{\rm D/H}|_{\rm P} - {\rm D/H}|_{\rm P}^{\rm Obs}\right]^2}{\sigma^2({\rm D/H}|_{\rm P} )^{\rm Theo} + \sigma^2({\rm D/H}|_{\rm P} )^{\rm Obs}} \, .
\end{align}
We are also interested in including a conservative determination of the baryon density to lift the degeneracy between $\Omega_{\mathrm{b}}h^2$ and $G$. Using directly the posterior values from the baseline $\Lambda$CDM Planck 2018 analysis will not be satisfactory, because $G$ is kept constant there. Instead, we use the results of \cite{Bai:2015vca}, who carry out a Planck likelihood analysis including variations in $G_{\mathrm{CMB}}$. In Table I of \cite{Bai:2015vca}, they find that the mean baryon density exactly matches that of the base Planck 2018 TTTEEE+lowE analysis within $\Lambda$CDM \cite{Aghanim:2018eyx}, albeit with twice as large error bars:
\begin{equation}
    \Omega_{\mathrm{b}}h^2|^{\mathrm{Obs}} = 0.02236, \quad \sigma(\Omega_{\mathrm{b}}h^2)= 0.00030.
\end{equation}
This allows us to define an extended $\chi^2$ for BBN+$\Omega_{\mathrm{b}}h^2$,
\begin{equation}
    \chi^2_{\mathrm{BBN}+\Omega_{\mathrm{b}}h^2} = \chi^2_{\mathrm{BBN}} + \frac{[\Omega_{\mathrm{b}}h^2 - \Omega_{\mathrm{b}}h^2|^{\mathrm{Obs}}]^2}{\sigma^2(\Omega_{\mathrm{b}}h^2)}\,.
\end{equation}
For both the pure BBN analysis and the extended scenario, we compute the relevant $\chi^2$ on a grid of $(\Omega_{\mathrm{b}} h^2,\,G_{\mathrm{BBN}}/G_0)$. We then marginalize over the baryon density to find a 1-D $\chi^2(G_{\mathrm{BBN}}/G_0)$. To rule out values of $G_{\mathrm{BBN}}/G_0$, we compare the computed statistics to critical values of the 1-D $\chi^2$ distribution. In particular, at 2$\sigma$, we rule out a scenario if $\Delta \chi^2 \equiv \chi^2 - \chi^2_\mathrm{min} \geq 4$.
\setlength\parskip{6pt}

\begin{figure}[t]
    \centering
    \includegraphics[width=\linewidth]{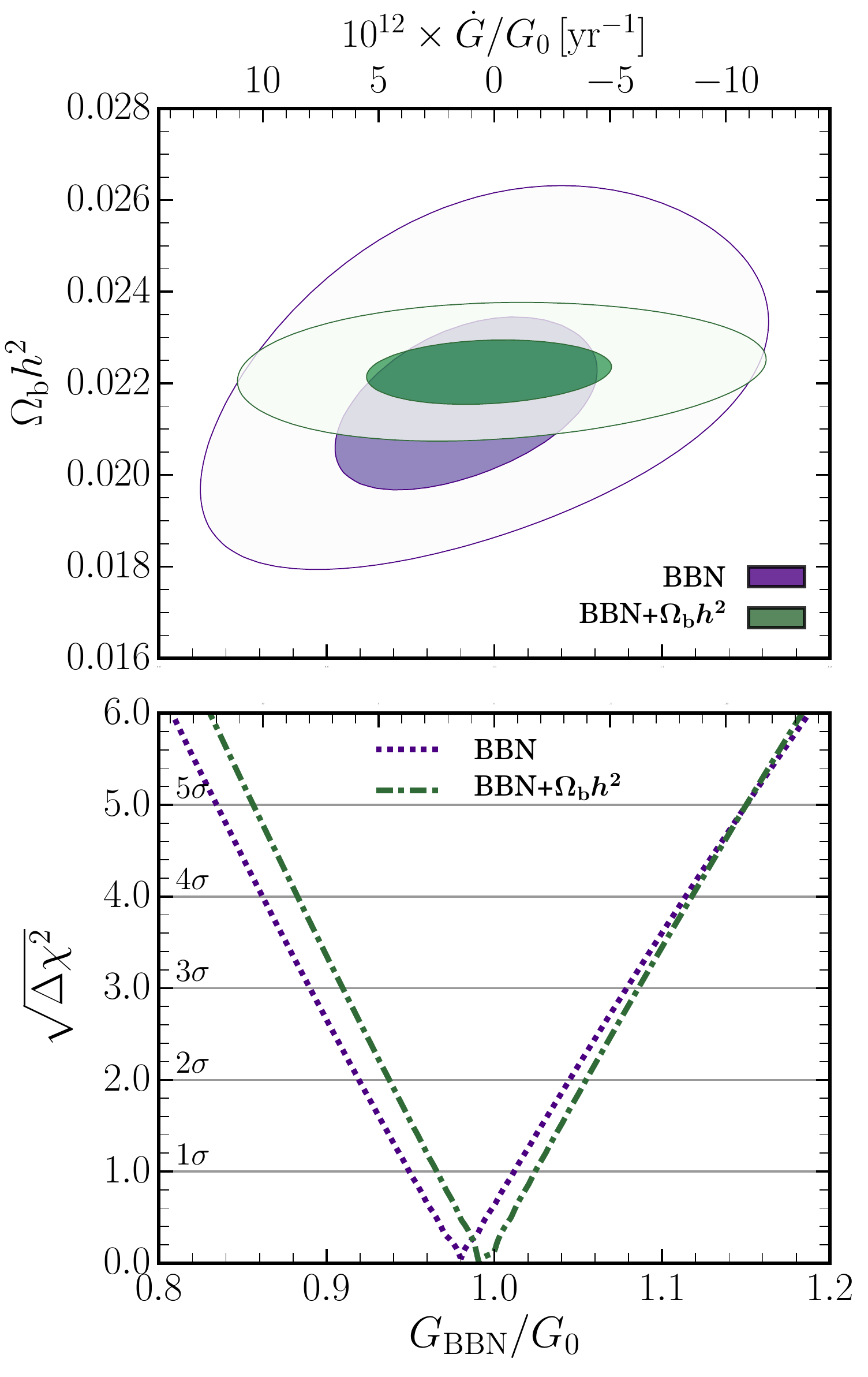}\vspace{-0.3cm}
    \caption{Contour plot showing the $1\sigma$ and $2\sigma$ confidence intervals in the $(\Omega_{\mathrm{b}}h^2, G_{\mathrm{BBN}}/G_0)$ plane (\emph{upper}) and marginalized $\Delta \chi^2$ as a function of $G_{\mathrm{BBN}}/G_0$ (\emph{lower}). The dotted lines correspond to BBN constraints and dash-dotted to BBN+$\Omega_{\mathrm{b}}h^2$ bounds.}
    \label{fig:bounds}
\end{figure}

\emph{Results.}--- We show the cosmological implications of a variation in the gravitational constant on the primordial helium and deuterium abundances in Figure \ref{fig:abundances}. It is evident that measurements of both primordial abundances are able to strongly constrain a deviation of $G$ from today's value. Moreover, we see that the impact of a higher value of $G_\mathrm{BBN}/G_0$ can be compensated by a higher value of the baryon density $\Omega_\mathrm{b}h^2$, which introduces a degeneracy in the $(\Omega_{\mathrm{b}}h^2, G_{\mathrm{BBN}}/G_0)$ plane.
This is because changes in the primordial deuterium abundance are linearly proportional to changes in the baryon density ($\Delta \mathrm{D}/\mathrm{H}|_\mathrm{P} \propto -\Delta\Omega_\mathrm{b}h^2$), while the primordial helium abundance is only logarithmically dependent on $\Omega_{\rm b} h^2$ \cite{Pitrou:2018cgg}.
\setlength\parskip{0pt}

In Figure \ref{fig:bounds} we show the $1\sigma$ and $2\sigma$ confidence intervals in the $(\Omega_{\mathrm{b}}h^2, G_{\mathrm{BBN}}/G_0)$ plane and $\Delta\chi^2$ as a function of the variation in $G$ for the two benchmark analyses considered here. At 95.4\% CL, we obtain:
\begin{align}
G_\mathrm{BBN}/G_0 &= 0.98^{+0.06}_{-0.06}\,  \qquad \text{BBN}, \\
G_\mathrm{BBN}/G_0 &= 0.99^{+0.06}_{-0.05}\,  \qquad\text{BBN+}\Omega_{\rm b}h^2.
\end{align}
Alternatively, assuming a linear time evolution of the gravitational constant as described in Eq.~\eqref{eq:gt}, these bounds can be translated into a constraint on $\dot{G}/G_0$, which at 95.4\% CL reads:
\begin{align}
\frac{\dot{G}}{G_0} &=  1.4^{+4.4}_{-4.7}\times 10^{-12}\,\, \mathrm{yr}^{-1} \quad\text{BBN}, \\
\frac{\dot{G}}{G_0} &= 0.7^{+3.8}_{-4.3}\times 10^{-12}\,\, \mathrm{yr}^{-1} \quad\hspace{0.2pt}\text{BBN+}\Omega_{\rm b}h^2.
\end{align}
Moreover, the BBN+$\Omega_\mathrm{b}h^2$ analysis highly disfavours a fractional deviation of $G$ larger than ${\sim}10\%$ (at more than $5\sigma$), while a similar conclusion can be drawn in the BBN-only analysis for deviations larger than ${\sim}20\%$. This difference arises because a restriction on $\Omega_\mathrm{b}h^2$ lifts the degeneracy with $G$ regarding the primordial deuterium abundance. Note that our constraints improve on the previous $1\sigma$ primordial nucleosynthesis bounds \cite{Copi:2003xd, Clifton:2005xr} by approximately a factor of 10.

As a final comment, other references, such as \cite{Pitrou:2018cgg}, quote an error on the deuterium value that is approximately $4$ times smaller than the one used here. Running our analysis with this lower error, we find no change in the BBN only bounds while the BBN+$\Omega_{\mathrm{b}}h^2$ constraints strengthen by ${\sim}30\%$.
\setlength\parskip{6pt}


\emph{Conclusions.}--- Big Bang Nucleosynthesis is sensitive to changes in the value of the gravitational constant in the early Universe. By using current measurements of the primordial abundances of helium and deuterium we have shown that at 95.4\% CL, $G_\mathrm{BBN}/G_0 = 0.98^{+0.06}_{-0.06}$ and $G_\mathrm{BBN}/G_0 = 0.99^{+0.06}_{-0.05}$ if $\Omega_{\rm b}h^2$ measurements from Planck are also accounted for. Assuming a very slow linear time evolution of $G$, these constraints map into a bound on the time variation at 95.4\% CL of $\dot{G}/G_0 = 1.4^{+4.4}_{-4.7}\times 10^{-12}\,\, \mathrm{yr}^{-1}$ and $\dot{G}/G_0 = 0.7^{+3.8}_{-4.3}\times 10^{-12}\,\, \mathrm{yr}^{-1}$ respectively. These constraints are competitive and complementary to those from CMB, stellar, solar system, pulsar timing, gravitational waves and lunar laser ranging measurements.
\setlength\parskip{6pt}

\emph{Acknowledgements.}--- We would like to thank A. Notari, G. Ballesteros and F. Rompineve for their very helpful correspondence on the first version of this manuscript. We acknowledge the use of the public cosmological code \texttt{PRIMAT}~\cite{Pitrou:2018cgg}. ME and MF are supported by the European Research Council under the European Union's Horizon 2020 program (ERC Grant Agreement No 648680 DARKHORIZONS). In addition, the work of MF was supported partly by the STFC Grant ST/P000258/1. JA is a recipient of an STFC quota studentship. NS is a recipient of a King's College London NMS Faculty Studentship.

\newpage

\bibliography{biblio}
\end{document}